\documentclass[twocolumn,showpacs,showkeys,preprintnumbers,amsmath,amssymb,prb]{revtex4}

\usepackage{graphicx}
\usepackage{dcolumn}
\usepackage{bm}

\begin{document}


\title{Spatially Resolved Raman Spectroscopy of Single- and Few-Layer Graphene}

\author{Davy Graf}
\email{grafdavy@phys.ethz.ch}
\author{Fran\c coise Molitor}
\author{Klaus Ensslin}
\affiliation{Solid State Physics Laboratory, ETH Zurich, 8093 Zurich, Switzerland}
\author{Christoph Stampfer}%
\author{Alain Jungen}%
\author{Christofer Hierold}%
\affiliation{Micro and Nanosystems, ETH Zurich, 8092 Zurich, Switzerland}%
\author{Ludger Wirtz}%
\affiliation{Institute for Electronics, Microelectronics, and Nanotechnology (IEMN), CNRS-UMR 8520, B.P. 60069, 59652 Villeneuve d'Ascq Cedex, France}%
\date{\today}

\begin{abstract}
We present Raman spectroscopy measurements on single- and few-layer graphene flakes. Using a scanning confocal approach we collect spectral data with spatial resolution, which allows us to directly compare Raman images with scanning force micrographs. Single-layer graphene can be distinguished from double- and few-layer by the width of the D' line: the single peak for single-layer graphene splits into different peaks for the double-layer. These findings are explained using the double-resonant Raman model based on {\it ab-initio} calculations of the electronic structure and of the phonon dispersion. We investigate the D line intensity and find no defects within the flake. A finite D line response originating from the edges can be attributed either to defects or to the breakdown of translational symmetry.
\end{abstract}

\pacs{Valid PACS appear here}
\maketitle


The interest in graphite has been revived in the last two decades with the advent of fullerenes \cite{Kroto85} and carbon nanotubes.\cite{Iijima91} 
However, only recently single- and few-layer graphene could be transferred to a substrate.\cite{Novoselov04} Transport measurements revealed a highly-tunable two-dimensional electron/hole gas of relativistic Dirac Fermions embedded in a solid-state environment.\cite{Novoselov05b, Zhang05} Going to few-layer graphene, however, disturbs this unique system in such a way that the usual parabolic energy dispersion is recovered.
The large structural anisotropy makes few-layer graphene therefore a promising candidate to study the rich physics at the crossover from bulk to purely two-dimensional systems. Turning on the weak interlayer coupling while stacking a second layer onto a graphene sheet leads to a branching of the electronic bands and the phonon dispersion at the K point. Double-resonant Raman scattering \cite{Thomsen00} which depends on electronic and vibrational properties turns out to be an ingenious tool to probe the lifting of that specific degeneracy.

We report on Raman mapping of single- and few-layer graphene flakes resting on a silicon oxide substrate. Lateral resolution of 400 nm allows to address neighboring sections with various layers of graphene down to a single graphene sheet, previously determined with the scanning force microscope (SFM). We find that the integrated G line signal is directly correlated with the thickness of the graphitic flake and is shifted upward in frequency for double- and single-layer graphene compared to bulk graphite. The mapping of the peak width of the D' line shows a strong contrast between single- and few-layer graphene. Such a pronounced sensitivity to the transition to the very last layer offers an optical and non-destructive method to unambiguously detect single-layer graphene. In addition, we locally resolve the structural quality of the flake by investigating the D band, which is related to elastic backscattering. The map of the integrated D line signal of a graphitic flake with double- and single-layer sections shows that the inner part of the flake is quasi defect free, whereas edges and steps serves as scatterers.
Finally, we explain the splitting of the D' line as a function of the number of 
graphene layers within the double-resonant Raman model.\cite{Thomsen00} The comparison between experimental data and 
theory confirms the qualitative validity of the double-resonant
Raman model, but demonstrates quantitative differences between
theory and experiment. 
In particular, the model, when based on first-principles
calculations, predicts a much smaller splitting of the peaks. 

\begin{figure}
\includegraphics[width=0.45\textwidth]{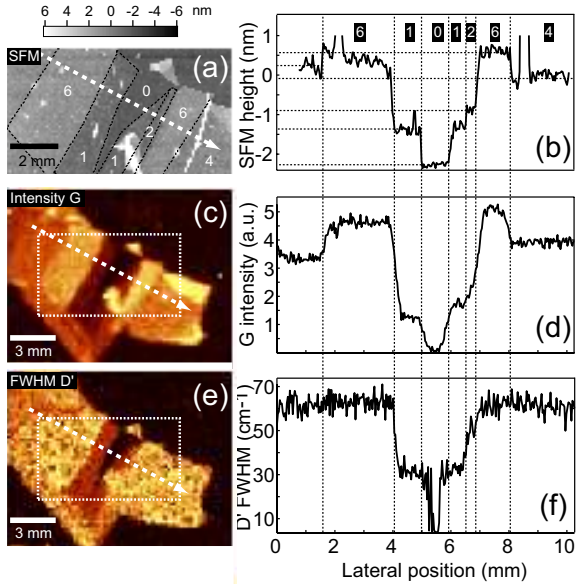}
\caption{\label{fig:ramanmap} (a),(b) SFM micrograph and cross-sectional plot (indicated with the white dashed arrow; lateral average over 400 nm) of a few-layer graphene flake with central sections down to a single layer. Raman maps (dashed square corresponding to the SFM image in (a)) showing (c) the integrated intensity of the G line and (e) the FWHM of the D' line. The related cross sections (d),(f) are aligned (vertical dashed lines) with the height trace.}
\end{figure}

The graphite films were prepared by mechanical exfoliation of highly oriented pyrolytic graphite (HOPG) and subsequent transfer to a highly doped Si waver with 300 nm SiO$_2$ (atomic oxidation process) cap layer. \cite{Novoselov04, Novoselov05} The combination of optical microscopy using phase contrast and SFM makes it possible to locate flakes with various thicknesses down to a monolayer with lateral extensions in the micrometer range.
The Raman spectra were acquired using a laser excitation of 532 nm (2.33 eV) delivered through a single-mode optical fiber, whose spot size is limited by diffraction. Using a long working distance focusing lens with a numerical aperture NA=0.80 we obtain a spot size of about 400 nm. With a very low incident power of 4-7 $\mu$W heating effects can be neglected.

The Raman spectrum of graphite has four prominent
peaks (Fig.~\ref{fig:ramanspectra} - for a recent review see Ref.~\onlinecite{Reich04}). The peak around 1582 
cm$^{-1}$, commonly called G line, is caused by the Raman active $E_{2g}$ phonon,
(in-plane optical mode) close to the $\Gamma$ point.
The D line around 1350 cm$^{-1}$ exhibits two remarkable features: its
position shifts to higher frequencies with increasing
incident laser excitation energies \cite{Vidano81} and its relative signal strength
(compared to the G line) depends strongly on the amount of disorder
in the graphitic material.\cite{Vidano81, Tuinstra70}
The associated overtone D' around 2700 cm$^{-1}$ is pronounced even in the
absence of a D signal. Finally, the overtone of the G line, the G' line,
is located at 3248 cm$^{-1}$, which is more than twice the energy of the
G line. The different experimental findings related to the
dispersive D, D' bands could be explained by Thomsen
and Reich within the framework of double resonant Raman scattering,\cite{Thomsen00} which was extended to other phonon branches by Saito \textit{et al.}\cite{Saito02} 
The electronic and vibrational properties of graphite are dominated by the sp$^2$-nature of the strong intraplane covalent bonds. The relatively weak inter-layer coupling causes the high structural anisotropy. 
Raman spectra for multiple graphene layers can be compared qualitatively and quantitatively while investigating flakes with sections of various thicknesses. In Fig.~\ref{fig:ramanmap}(a) the SFM micrograph of a graphite flake with different layers is presented: The bare SiO$_{2}$ (indicated by '0') is surrounded by single-layer sections with steps of up to two, six and four layers. The different step heights are clearly depicted in Fig.~\ref{fig:ramanmap}(b), where a cross section of Fig.~\ref{fig:ramanmap}(a) (see white dashed arrow) is shown. Scanning the flake and collecting for each spot the complete Raman spectrum we can subsequently filter specific spectral data for spatially resolved data point and construct false-color 2D plots. In Fig. \ref{fig:ramanmap}(c) the intensity of the G peak is integrated from 1537 to 1622 cm$^{-1}$. We find a remarkable correlation with the SFM graph: Brighter regions correspond to thicker sections. The cross section in Fig. \ref{fig:ramanmap}(d) shows a step-like behavior, perfectly correlated with the topographical changes shown in Fig.~\ref{fig:ramanmap}(b).
In Fig. \ref{fig:ramanmap}(e) we plot the FWHM (full width at half maximum) of the D' line. It shows the narrowing at the transition to a single-layer (see e.g. Fig.~\ref{fig:ramanspectra}) and gives an evident contrast between single- and few-layer graphene sections. The quasi digital change from about 60 to 30 cm$^{-1}$ shown in Fig.~\ref{fig:ramanmap}(f) suggests that the width of the D' line can be used as a detector for single-layer graphene resting on a substrate. 
Raman spectroscopy can therefore be used to count the layers of a thin graphite stack and to discriminate between single and double layer. 
Combined with the double-resonant Raman scattering mechanism an optical setup using light in the visible range 
turns out to be an alternative to scanning force microscopy, which requires stacking folds as height references.

\begin{figure}
\includegraphics[width=0.45\textwidth]{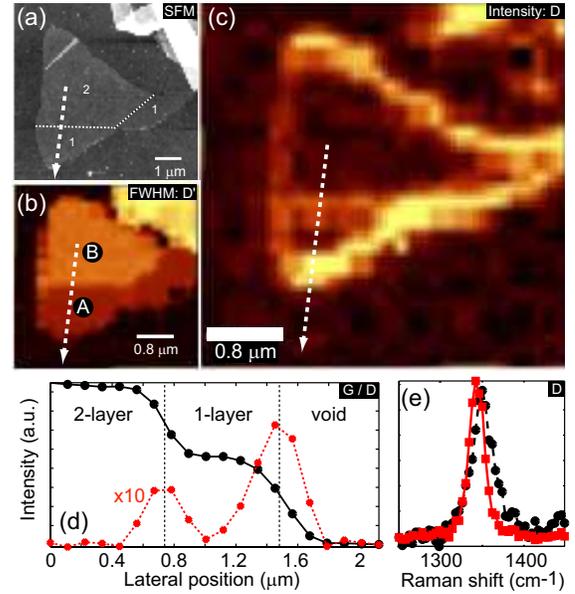}
\caption{\label{fig:ramandline} (a) SFM micrograph of a graphitic flake consisting of one double- and two single-layer sections (white dashed line along the boundaries), highlighted in the Raman map (b) showing the FWHM of the D' line.  (c) Raman mapping of the integrated intensity of the D line: A strong signal is detected along the edge of the flake and at the steps from double- to single-layer sections. (d) Raman cross section (white dashed arrow in (c)): Staircase behavior of the integrated intensity of the G peak (solid line) and pronounced peaks at the steps for the integrated intensity of the D line (dashed line). (e) Spatially averaged D peak for the crossover from double to single layer ($\bullet$, dashed line) and from single layer to the SiO$_2$ substrate ($\blacksquare$, solid line).}
\end{figure}

Transport measurements show that the quality of the finite graphitic flakes on the silicon oxide matrix obtained with the technique explained above is remarkable: electronic mobilities up to 15'000 cm$^2$(Vs)$^{-1}$ were estimated from transport experiments. \cite{Novoselov05b, Zhang05}
We point out that also the Raman spectroscopy reveals quasi defect free graphitic sheets via the absence of a D band signal.
First experiments have related the intensity of the D band to the structural coherence of the graphite material. In fact it is inversely proportional to the crystallite grain size. \cite{Tuinstra70} The appearance of the D band can, however, be related to the occurrence of defects and disorder in general, as shown in experiments with boron-doped and electrochemically oxidized HOPG.\cite{Wang90} With micro-Raman mapping we are able to localize the spatial origin of the defects. From cross-correlating the SFM micrograph in Fig.~\ref{fig:ramandline}(a) with the Raman map of the integrated D line (1300-1383 cm$^{-1}$) intensity in Fig. \ref{fig:ramandline}(c) we infer directly that the edges of the flake and also the borderline between sections of different height contribute to the D band signal whereas the inner parts of the flakes do not. This is somewhat surprising since for thinner flakes the influence of a nearby substrate on the structural quality should be increasingly important.
In the cross-section Fig.~\ref{fig:ramandline}(d) we see clearly that the D line intensity is maximal at the section boundaries, which can be assigned to translational symmetry breaking or to defects. However, we want to emphasize that the D line is still one order of magnitude smaller than the G line.
In Fig. \ref{fig:ramandline}(e) spatially averaged D mode spectra from the two steps shown in Fig.~\ref{fig:ramandline}(d) are presented. The frequency fits well into the linear dispersion relation of peak shift and excitation energy found in earlier experiments.\cite{Vidano81} In addition, we find that the peak is narrower and down-shifted at the edge of the single-layer while it is somewhat broader and displays a shoulder at the crossover from the double to the single layer.

\begin{figure}
\includegraphics[width=0.45\textwidth]{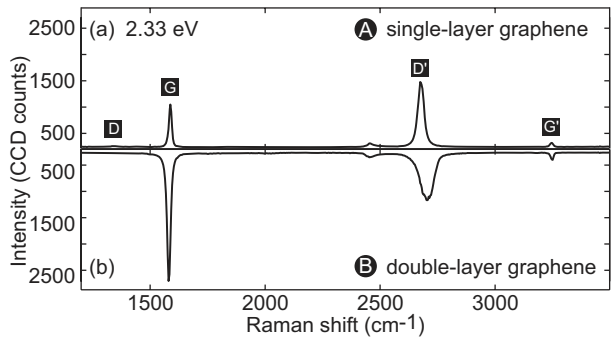}
\caption{\label{fig:ramanspectra} Raman spectra of (a) single- and (b) double-layer graphene (collected at spots A and B, see Fig. \ref{fig:ramandline}(b)).}
\end{figure}

\begin{figure}
\includegraphics[width=0.45\textwidth]{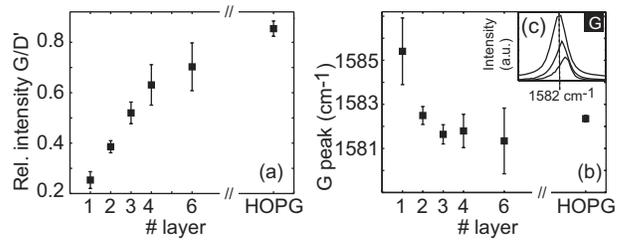}
\caption{\label{fig:ramanaanalysis} (a) Plot of the ratio of the integrated intensities of the G and D' peak versus number of stacked layers (average value and standard deviation). (b) G line frequency versus number of stacked layers (average value and standard deviation). (c) G peak for HOPG (upper peak), double- (middle peak) and single-layer (lower peak) graphene. The vertical dashed line indicates the reference value for bulk graphite.}
\end{figure}

In Fig.~\ref{fig:ramanspectra}(a) and (b), we compare the Raman spectra of the double- and single-layer graphene shown in Fig.~\ref{fig:ramandline}(b) 
and labeled with A and B. The Raman signal is significantly altered when peeling off the penultimate layer: the G peak decreases strongly in intensity and shifts towards higher wave numbers. 
In connection with Fig.~\ref{fig:ramanmap}(b) we already stated that the integrated G line signal is monotonously increasing with increasing flake thickness. In order to compare data of different flakes and measurement runs we turn our attention to the ratio of the integrated intensities of the G and D' line, plotted in Fig. \ref{fig:ramanaanalysis}(a). Most of the changes can be attributed to the decrease of the G line, since the spectral weight of the D' band changes only slightly. The intensity ratio increases almost linearly from one to four layers.
In Fig. \ref{fig:ramanaanalysis}(b) the dependence of the G peak position on the layer number is investigated. Spectral data of various sections on different flakes were averaged. The frequency shifts towards higher wave numbers at the crossover to the double- and especially to the  single-layer graphene. However, in the case of single-layer graphene, it is accompanied by an important statistical spread of the collected data. 
In Fig. \ref{fig:ramanaanalysis}(c) representative G peak spectra for single-, double-layer graphene and HOPG are presented. 
It is important to note that in contrast to the G line, the corresponding overtone band, the G' line, does not change its spectral position as a function of the number of layers.

\begin{figure}
\includegraphics[width=0.3\textwidth]{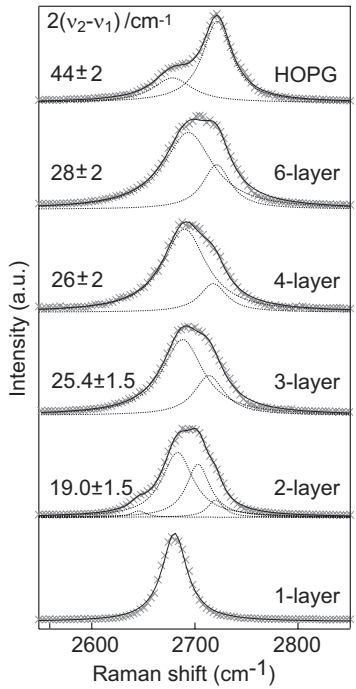}
\caption{\label{fig:ramanaanalysisDprime} D' peaks for an increasing number of graphene layers along with HOPG as a bulk reference. The dashed lines show the Lorentzian peaks used to fit the data, the solid lines are the fitted results. The single peak position for the single-layer graphene is at 2678.8 $\pm$ 1.0 cm$^{-1}$. The peak position of the the two inner most peaks for double-layer graphene are 2683.0 $\pm$ 1.5 and 2701.8 $\pm$ 1.0 cm$^{-1}$. On the left the value for the splitting from double-layer graphene up to HOPG is presented. All peaks are normalized in amplitude and vertically offset.}
\end{figure}

\begin{figure}[hbtp]
 \centering
   \includegraphics[draft=false,keepaspectratio=true,clip,%
                    width=0.9\linewidth]%
                   {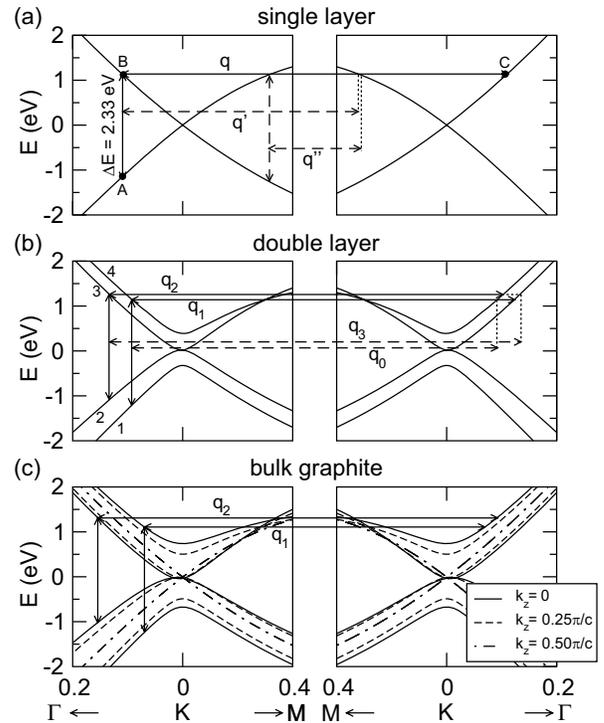}
\caption{Electronic band structure along the high-symmetry lines
$\Gamma$-K and K-M: (a) single-layer graphene, (b) double-layer
graphene, and (c) bulk graphite. For bulk graphite, we 
display the band structure in the direction parallel to 
the graphene planes for different values of the transverse momentum $k_z$.
Vertical arrows denote vertical transitions by 2.33 eV from a valence
($\pi$) band to a conductance ($\pi^*$) band. Horizontal arrows denote
transitions between two states of almost equal energy by coupling
to a phonon of momentum $q_i$ (the corresponding phonon frequencies
are displayed in Table \ref{tab:freqtable}).
Dashed horizontal lines denote transitions
with considerably less weight than the solid horizontal lines (see text).}
\label{fig:bandstrucs} 
\end{figure}

The most prominent difference in the spectra of single-layer, 
few-layer, and bulk graphite lies in the D' line:
the integrated intensity of the D' line stays almost constant, even though it narrows to a single peak at lower wave number at the crossover to a single layer (Fig.~\ref{fig:ramanspectra}). 
The width of the D' peak or - at high resolution - its splitting into different sub-peaks (Fig.~\ref{fig:ramanaanalysisDprime})
is in the following explained in the framework of the 
double-resonant Raman model. \cite{Thomsen00}
The model explains the D' line in the following way (see Fig.~\ref{fig:bandstrucs}(a)): An electron is vertically excited from point A in the $\pi$ band
to point B in the $\pi^*$ band by absorbing a photon. The excited electron
is inelastically scattered to point C by emission of a phonon with momentum $q$.
Since the energy of this phonon ($\approx 150$ meV) is small compared
with the photon energy of 2.33 eV, we have drawn the line horizontally,
for simplicity. Inelastic backscattering to the vicinity of point A by
emission of another phonon with momentum $\approx q$ and
electron-hole recombination lead to emission of a photon with an energy
about 300 meV less than the energy of the incident photon.
In principle, two other double-resonant Raman processes, involving
the phonons $q'$ and $q''$, are possible as well. However, it was argued
in Ref.~\onlinecite{Ferrari06} that their weight is very low.\cite{lowweight}

In Fig.~\ref{fig:bandstrucs}, we compare the electronic band structure
of the single layer with the ones of the double layer and of bulk graphite.
All three dispersion relations were calculated from first-principles,
using density-functional theory in the local density approximation.\cite{abinit}
In the double-layer, the $\pi$ and $\pi^*$ bands split into two bands
each. This gives rise to four different possible excitations.
We have calculated the corresponding oscillator strengths \cite{self}
and found that for the excitation energy of 2.33 eV, transitions 1--3
and 2--4 have negligible weight, while transitions 1--4 and 2--3
(displayed in Fig.~\ref{fig:bandstrucs}(b)) have almost equal weight.
For each of the two dominant vertical transitions, there are two 
possible horizontal transitions.
The corresponding electron-phonon coupling matrix elements for the 
phonons $q_0$ to $q_3$ are almost equal.\cite{lazzeri}
In theory, we therefore expect a splitting of the D' band into four
peaks of almost equal height. Our experimental data (Fig.~\ref{fig:ramanaanalysisDprime})
shows indeed that the D' line for the double layer can be 
decomposed into four peaks. However, the outer two peaks
(corresponding to the phonons $q_0$ and $q_3$) have very 
low weight in the experimental data. We calculated from first principles
\cite{abinit} the phonon-frequencies $\nu_1$ and $\nu_2$, corresponding to the momenta
$q_1$ and $q_2$. The frequencies of the highest optical branch are
given in Table~\ref{tab:freqtable}. Due to the weak inter-layer coupling
the degeneracy of this branch is lifted. However, the frequency difference
remains weak ($<$ 1 cm$^{-1}$) and does not significantly contribute to
the experimentally observed splitting about 19 cm$^{-1}$ of the D' line (see Fig.~\ref{fig:ramanaanalysisDprime}).
Table~\ref{tab:freqtable} furthermore gives the value for $2(\nu_2-\nu_1)$.
We note that the value obtained from our first-principles
calculation is only half as large as the experimentally observed
splitting of about 19 cm$^{-1}$. This discrepancy is related
to the fact, that the double-resonant Raman model based on
\textit{ab-initio} calculations also predicts a value for the
dispersion of the D' line with incident laser energy
that amounts only to about half of the experimentally observed
value of 99 cm$^{-1}$/eV.\cite{compnote}
We conclude therefore that the double-resonant Raman
model can qualitatively explain the fourfold splitting of the D' line
in the double-layer, but the amount of the splitting and the relative
heights of the peaks are not properly described within this model.\cite{newnote}

In bulk graphite, the $\pi$ and $\pi^*$ bands split into a continuum
of bands, i.e., they disperse in the direction $k_z$ perpendicular to
the layer. In Fig.~\ref{fig:bandstrucs}(c), we display the bands for three
different values of $k_z$. In the joint-density of states, the
vertical transitions for $k_z=0$ have the dominant weight and are thus
considered in our calculations. Since the splitting between the bands
is much more pronounced than for the double-layer, the value for
$2(\nu_2-\nu_1)$ is about a factor of two higher than for the 
double layer. 
This is in agreement with the experimental data, see Fig.~\ref{fig:ramanaanalysisDprime}, where the
splitting increases likewise by about a factor of two between the
double layer and bulk graphite. As in the case of the double layer, there are
quantitative differences between theory and experiment for graphite as well: First-principles
calculations of the oscillator strengths and of electron-phonon coupling
matrix elements predict an almost equal height of the peaks whereas
the experiment shows that the lower-frequency peak has a strongly 
reduced weight. The peaks corresponding to the horizontal 
transitions $q_0$ and $q_3$ are missing altogether in the experimental
spectrum.

\begin{table}
\begin{tabular}{|l|c|c||c|}
\hline
 & $\nu_1$/cm$^{-1}$ & $\nu_2$/cm$^{-1}$ & $2(\nu_2-\nu_1)$/cm$^{-1}$ \\
\hline
\hline
bulk & 1393.2/1393.6 & 1402.9/1403.1 & 19.4/19.0\\
\hline
double layer & 1395.6/1395.6 & 1400.0/1400.6 & 8.8/10.0\\
\hline
single layer & \multicolumn{2}{|c||}{1398.1}  & - \\
\hline
\end{tabular}
\caption{Frequencies of the optical phonons involved in the double-resonant
Raman model. The corresponding phonon momenta $q_1$ and $q_2$ are 
determined from the {\it ab-initio} electronic band structures of Fig.~\ref{fig:bandstrucs}.
The splitting of the frequencies in the double-layer and bulk
is due to the (weak) inter-layer interaction.}
\label{tab:freqtable}
\end{table}


Even though some quantitative differences remain, the double-resonant
Raman model explains well the observed differences in the D' line
as we go from the single-layer via few-layer systems to the bulk limit.
The quantitative differences may be an indicator that some essential
effects are not properly included in the model. E.g., the role of quasi-particle
effects (electron-electron interaction)\cite{gwnote} and of excitonic effects
(electron-hole interaction) in the double-resonance process
remains to be understood. 
The importance of these effects has been recently 
demonstrated for electronic excitations in carbon nanotubes (both 
semiconducting and metallic).\cite{spataru, chang} A similar importance 
may be therefore expected for processes that involve electronic excitations 
in graphite.\cite{tbnote}

In conclusion, Raman mapping reveals to be a powerful tool to investigate single- and few-layer graphene flakes. 
It turns out that the width of the D' line is highly sensitive to the crossover from single- to double-layer graphene, which is explained by a peak splitting following the double-resonant Raman model together with \textit{ab initio} electronic band structure calculations.
A remaining open question is the decrease of the G line intensity with decreasing layer number compared to the almost constant spectral weight of the D' line and the accompanied upshift of its frequency for double- and single-layer graphene. 
The structural quality of the flakes is studied by analyzing the D line intensity: no defects are detected in the inner part of the flake. The D line signal from the boundaries of the individual sections of the flake suggest that they act as elastic scatterer.

The authors are grateful to Hubert Heersche for useful advices on sample preparation. 
We acknowledge stimulating discussions with A. Rubio and M. Lazzeri.
Calculations were performed at IDRIS (project 061827). Financial support from the Swiss Science Foundation (Schweizerischer Nationalfonds) is gratefully acknowledged. L. W. acknowledges support from the French National Research Agency.


\begin{thebibliography}{99}
\bibitem{lowweight} The process involving $q'$
(a phonon very close to the K-point) does not
occur because the corresponding electron-phonon coupling matrix element
is approximately 0 (see footnote [24] of Ref.~\onlinecite{piscanec}).
And the process involving $q''$ has a low weight because
due to the  ``trigonal warping'' effect, the corresponding phase space
volume in the two-dimensional Brillouin
zone is small compared to the phase space volume for process involving 
$q$.
\bibitem{newnote}
For an exact quantitative evaluation of the peak heights, an integration
in the two-dimensional Brillouin zone for all transitions of energy
$\Delta E = 2.33$ eV would have to be performed.
\bibitem{tbnote} We note that the picture of independent 
electron-hole pair excitations has 
been quantitatively very successful for the assignment of
nanotube spectra to the chiral indices (m,n).
However, this success was based on the use of tight-binding
band-structures with correspondingly fitted parameters.
A parameter-free, first-principles, calculation of the optical spectra of
carbon nanotubes must take into account electron-electron and
electron-hole interaction, and we suggest that this may be
true for graphite as well.
\end{thebibliography}
\end{document}